\documentclass[proof]{pasj00}
\usepackage{graphicx}
\usepackage{comment}
\usepackage{color}
\usepackage{pst-grad}
\usepackage{psfrag}
\usepackage{url}
      
\begin{document}
\SetRunningHead{Astronomical Society of Japan}{}
\newcommand{\sun}{\ensuremath{\odot}}%
\newcommand{\vdag}{(v)^\dagger}
\newcommand{\Gray}{$\gamma$-ray~}
\newcommand{\Grays}{$\gamma$-rays~}
\newcommand{\mf}{100~$\mu$G~}
\newcommand{\hii}{H{\sc ii} region}


\title{Modeling the gamma-ray emission produced by runaway cosmic rays in the environment of RX~J1713.7-3946}

\author{Sabrina \textsc{Casanova} \altaffilmark{1},
          \email{sabrina.casanova@mpi-hd.mpg.de}
        David I. \textsc{Jones} \altaffilmark{1},
        Felix A.  \textsc{Aharonian}\altaffilmark{1,2},
        Yasuo  \textsc{Fukui}\altaffilmark{3},
        Stefano \textsc{Gabici}\altaffilmark{2,4},
        Akiko \textsc{Kawamura}\altaffilmark{3},
        Toshikazu \textsc{Onishi}\altaffilmark{3},
        Gavin \textsc{Rowell}\altaffilmark{5},
        Hidetoshi \textsc{Sano} \altaffilmark{3},
        Kazufumi \textsc{Torii} \altaffilmark{3},
        Hiroaki \textsc{Yamamoto}\altaffilmark{3}
}

\altaffiltext{1}{Max Planck f\"ur Kernphysik, Saupfercheckweg 1, 69117, Heidelberg}
\altaffiltext{2}{Dublin Institute for Advance Physics,31 Fitzwilliam Place, Dublin 2, Ireland}
\altaffiltext{3}{Nagoya University,Furo-cho, Chikusa-ku, Nagoya City, Aichi Prefecture, Japan}
\altaffiltext{4}{Laboratoire APC, 10, rue Alice Domon et L\'{e}onie Duquet, 75013 Paris, France}
\altaffiltext{5}{School of Chemistry and Physics, University of Adelaide, Adelaide 5005, Australia}




\maketitle

\begin{abstract}

Diffusive shock acceleration in supernova remnants is the 
most widely invoked paradigm to explain the Galactic cosmic ray spectrum. 
Cosmic rays escaping supernova remnants diffuse in the 
interstellar medium and collide with the ambient atomic and molecular gas. 
From such collisions gamma-rays are created, which can possibly provide 
the first evidence of a parent population of runaway cosmic rays. 
We present model predictions for the GeV to TeV gamma-ray emission 
produced by the collisions of runaway cosmic rays with the gas 
in the environment surrounding the shell-type supernova remnant RX~J1713.7-3946. 
The spectral and spatial distributions of the emission, which depend upon 
the source age, the source injection history, the diffusion regime and the 
distribution of the ambient gas, as mapped by the LAB and NANTEN surveys, 
are studied in detail. In particular, we find for the region surrounding 
RX~J1713-3946, that depending on the energy one is observing at, 
one may observe startlingly different spectra or may not detect any enhanced 
emission with respect to the diffuse emission contributed by background cosmic rays. 
This result has important implications for current and future gamma-ray experiments.
\end{abstract}


\section{Introduction}\label{sec:intro}

Cosmic rays (CRs) are the highly energetic protons and nuclei which fill 
the Galaxy and carry, at least 
in the vicinity of the Sun, as much energy per unit volume as the energy 
density of starlight or 
of the interstellar magnetic fields or the kinetic energy density of the interstellar gas. 

CRs of energies up to the ``knee'' (${10}^{15}$ eV), or even up to 
${10}^{18}$ eV, are believed to originate from sources located within the 
Galaxy. Galactic CRs are thought to be
accelerated via diffusive shock
acceleration (DSA) operating in the expanding shells of supernova remnants (SNRs). For 
references see e.g. \cite{Blandford,Malkov}.
The accelerated CRs interact with ambient gas through
inelastic collisions, producing neutral pions, which then decay into \Grays. 

If the bulk of Galactic CRs up to at least PeV energies are indeed 
accelerated in SNRs, then TeV \Grays are expected to be emitted during 
the acceleration process CRs undergo within SNRs \citep{Drury}. Indeed 
TeV \Grays have been detected from the shells of SNRs \citep{Aharonian:nature,Albert}. However, such 
observations do not constitute a definitive proof that CRs are 
accelerated in SNRs, since the observed
emission could be produced by energetic electrons up scattering low energy photon fields. We note, 
in fact, that \Grays are also produced through inverse Compton and 
bremsstrahlung processes of very highly energetic electrons.

\Grays are also expected to be emitted when the accelerated 
CRs propagate into the interstellar medium (ISM) \citep{Montmerle,Issa,Aharonian:1991,Aharonian:1996,Gabici2009}. 
Before being isotropised by the Galactic magnetic fields, the injected CRs produce $\gamma$-ray emission, 
which can significantly differ from the emission of the SNR itself, as 
well as from the diffuse emission contributed by the background CRs and  electrons, 
because of the hardness of the runaway CR spectrum, which is not 
yet steepened by diffusion. The extension of such diffuse sources does not 
generally exceed a few hundred parsecs, 
the scale at which the spectra of the injected CRs can significantly 
differ from the spectrum of the CR background \citep{Aharonian:1996,Gabici,Marrero,Gabici2009}. 
These diffuse sources are often correlated with dense molecular clouds (MCs), which 
act as a target for the production of \Grays due to the enhanced local CR injection spectrum. \cite{Casse} 
and \cite{Montmerle} have pointed out that SNRs are located in star forming regions, 
which are rich in molecular hydrogen. In other words, CR sources and MCs 
are often associated and target-accelerator systems are not 
unusual within the Milky Way. In fact, there have been recently claims of detection of \Grays in association with dense MCs close to 
candidate CR sources, both at GeV energies \citep{Abdo2009d,Abdo2010,Tavani,Castro} and TeV energies 
\citep{Albert2007,Aharonian:2008,Aharonian:2008b,Aharonian:2008c}. 

In order to probe the interaction of the low-energy component of the 
ambient cosmic ray flux, \cite{Montmerle:2009} has recently 
proposed to measure enhanced ionization in TeV-bright molecular 
clouds, using millimeter observations. On the other hand, high energy CRs (above 1 GeV up to PeV energies) 
interacting with the ambient gas emit \Grays. For this reason the 
high sensitivity, high resolution \Gray data from current (HESS, Magic, Veritas, Fermi and Agile) 
and future detectors, such as AGIS, CTA and HAWC \citep{Sinnis,Hinton}, together with the knowledge 
of the distribution of the atomic and molecular hydrogen in the Galaxy on sub-degree scales 
are crucial to explore the flux of high energy CRs close to the candidate CR sources and to pinpoint the long searched-for 
sites of CR acceleration.

The \Gray radiation from hadronic interactions from regions close to CR sources depends not only on the total
power emitted in CRs by the sources and on the distance 
of the source to us, but also on the ambient interstellar
gas density, the local diffusion coefficient and the injection history of the CR source.
It is therefore difficult to definitely recognize the sites of CR acceleration from 
\Gray observations alone, since very often only qualitative predictions 
are provided, rather than robust quantitative predictions, especially
from a morphological point of view. In order to fully exploit the present and future experimental
facilities and to test the standard scenario for CR injection in SNRs and propagation, 
we present here model predictions of the spectral and morphological
features of the hadronic \Grays emission surrounding the candidate CR source, RX~J1713.7-3946, 
by constructing as quantitative a model as possible. In particular, we will convey 
all information concerning the environment, the source age, 
the acceleration rate and history, which all play a role in the physical 
process of CR injection and propagation. Building upon the modeling of the broadband emission from MCs close to CR 
accelerators developed in \cite{Gabici2009} and upon the analysis of the CR background discussed 
in \cite{Casanova2009} (hereafter Paper 1), we compute the expected \Gray emissivity from hadronic interactions 
of runaway CRs for the region $340^\circ<l<350^\circ$ and $-5^\circ<b<5^\circ$, assuming 
that a historical SNR event occurred in 393 C.E., at the location of the SNR RX~J1713.7-3946 \citep{Wang}.

RX~J1713-3946 is thought of as one of the best examples of a shell-type 
SNR, for which the multi-wavelength data suggests hadronic CR particle acceleration 
is active up to at least 100 TeVs \citep{Aharonian:nature}. The acceleration 
site within RX~J1713.7-3946 is spatially coincident with the sites of 
non-thermal X-ray emission and brightening and decay of the X-ray hot spots on year time-scales 
have been detected \citep{Uchiyama}. The observed rapid variability of the X-ray emission provides strong 
evidence for the amplification of the magnetic field around the SNR shell, which is 
is key condition for the acceleration of protons 
beyond the 100 TeV limit set by \cite{Lagage}. 
The multi-wavelength analysis of the emission from RX~J1713.7-3946 supports therefore 
the hypothesis that CRs up to the knee are accelerated in SNRs. 
However, no compelling evidence for the acceleration of protons and nuclei up to PeV energies 
has been found until now. Also, the most energetic 
CRs cannot be confined for long time within the SNR and, even if RX~J1713-3946 might have accelerated 
particles up to about PeV energies once, such highly energetic protons have already left the source. In fact, these 
very energetic particles, which are released first by SNRs and diffuse faster than lower energy CRs, 
reach as first the clouds surrounding the injection sites and produce 
enhanced \Gray emission. This is why studies of the \Gray emission from environment 
surrounding RX~J1713-3946, such as the one presented here, are important.

In Section \ref{sec:surveys} we will briefly describe the results 
of the LAB survey of the Galactic atomic hydrogen 
and of the NANTEN survey of the Galactic molecular clouds. Section
\ref{sec:model} will be dedicated to the
description of the model. The predictions of the model for the emission from the
region under consideration are presented in Section \ref{sec:results},
where we will also discuss the observational prospects for present and
future observatories.
Our conclusions are given in Section \ref{sec:conclusions}.

\section{Survey data}\label{sec:surveys}

The distributions of atomic and molecular hydrogen in the Galaxy, used in the
following calculations, are from the Leiden/Argentine/Bonn (LAB) Galactic
HI Survey \citep{Kalberla}
and from the NANTEN Survey \citep{Fukui1,Fukui2,Nanten,Fukui3}, respectively.

The LAB observations of atomic hydrogen were centered on the
1420~MHz ($\lambda21$~cm) line with a bandwidth of 5~MHz.  The
velocity axis spans $-450$~km~s$^{-1}$ to $400$~km~s$^{-1}$ giving a
final velocity resolution of $1.3$~km~s$^{-1}$.  The data combines
observations from three telescopes at a resolution of $0.6^\circ$
with a final sensitivity of $\sim0.1$~K.

The NANTEN instrument is a 4~m millimetre/sub-millimetre telescope, which
has surveyed the southern sky,
using the  $^{12}$CO (J=1--0) emission line at 115.271~GHz
($\lambda=2.6$~mm). The survey was
performed with an angular resolution of 4$'$, with a mass sensitivity of
about
100~M$_\odot$ at the Galactic centre and 1~km/s resolution in velocity. From the column density of
CO the column density of molecular hydrogen is obtained by assuming the
conversion factor, $X$,
 \begin{equation}\label{eq:X}
 X= 1.4 \times {10}^{20}  \, e^{(R/11 \, {\rm kpc})}{\rm {~[cm}^{-2}
{K}^{-1} {km}^{-1} s]},
 \end{equation}
where $R$ is distance to the GC \citep{Nakanishi2003,Nakanishi2006}.

\begin{figure}
  \begin{centering}
\includegraphics[width=0.45\textwidth,angle=0]{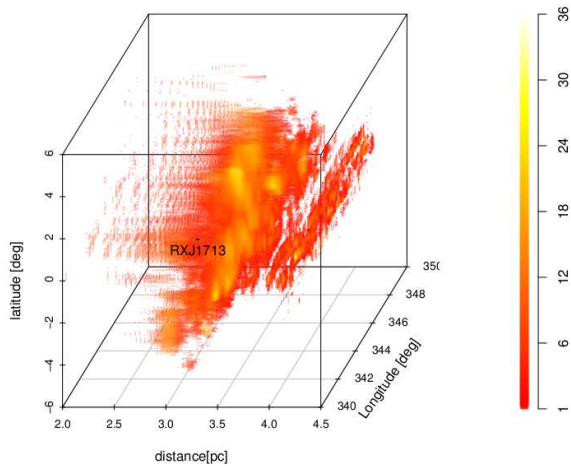}
\caption{The gas distribution in the region which spans Galactic longitude
$340^\circ<l<350^\circ$,
Galactic latitude $-5^\circ<b<5^\circ$ and heliocentric distance 100 pc $<l_d<$ 30 kpc,
as observed by the NANTEN and LAB surveys, expressed in protons cm$^{-3}$. 
The distance axis is logaritmic in base 10.
A value for the gas density
is given every 50 pc in distance, which is reflected in the apparent slicy structure 
for distances below 100 pc. For sake of clarity 
only densities above 1 protons cm$^{-3}$ are shown. 
Also indicated the position of the
historical SNR, RX~J1713.7-3946.\label{fig0}}
 \end{centering}
\end{figure}

The survey data-cubes of both atomic and molecular 
hydrogen, given in longitude, latitude and velocity, are transformed into 
data-cubes in longitude, 
latitude and heliocentric distance by using a flat rotation curve model of 
the Galaxy with uniform velocity equal to $220 \, {\rm km/s}$ 
\citep{Nakanishi2003,Nakanishi2006}. In Figure \ref{fig0} 
the three dimensional gas distribution (sum of the atomic and molecular gas) 
in the region which spans Galactic longitude $340^\circ<l<350^\circ$, 
Galactic latitude $-5^\circ<b<5^\circ$ and distance {\bf 100} pc $<l_d<$ 30 kpc 
is shown. For a detailed discussion of the data and its limitations and errors, 
we refer the reader to Paper I.

\section{The model}\label{sec:model}
We have modeled a supernova (SN) of total energy $E_{SN}$, which has exploded into the 
ISM in the region of the Galaxy described above. Our SNR is identifiable with the
historical SNR RX~J1713.7-3946, is located at 1 kpc distance from the Sun and has the same coordinates 
as RX~J1713.7-3946, i.e., (l,b) = (347.3,-0.5) \citep{Wang}. CRs are assumed to 
be injected by DSA operating at SNR shocks and the spectrum at the shock is a power law with index -2. 
The SNR accelerates the most energetic particles at the transition from 
the free expansion phase to the Sedov phase \citep{Ptuskin,Gabici2009,Caprioli}. Following 
\cite{Gabici2009} we assume that the most energetic leave the SNR at the beginning of the 
Sedov phase. The particles 
are released at different times, depending on their energy. A particle with 
energy $E_p$ escapes the SNR at a time $\chi(E_p)$, for which, following \cite{Gabici2009},
we assume a power law behaviour 
\begin{equation}
\chi(E_p) = t_{Sedov} \, {(\frac{E_p}{{E_{p}}_{max}})}^{-1/\delta} \,.
\label{eqn:chi}
\end{equation}
The maximum injection energy in Eq. \ref{eqn:chi}
is assumed to be ${{E_{p}}_{max}}$ = 500 TeV. If we assume that CRs of energy ${E_{p}}_{max}$ and ${E_{p}}=150$ TeV are released at an early epoch of the Sedov phase (t= 100 yr) and now (t= 1600 yr), 
respectively, then this requires $\delta=0.43$. When runaway protons 
diffuse into the ISM,
their energy density varies with energy, $E_p$, distance from the
injection source, $d$, SNR age, $t$, and diffusion properties of the ISM \citep{Gabici2009} as
\begin{eqnarray*}
\rho_{SNR}(E_{p},d,t) &=&  \frac{\eta  \, E_{SN}}{{\pi}^{3/2} \,ln({E_{p}}_{max}/{E_{p}}_{min})}\\
&& \times \, \frac{e^{-{(\frac{d}{R_d})}^{2}}}{{R_d}^3}  \, {E_p}^{-2} \,.
\end{eqnarray*}
\begin{equation}
\label{equCRdensitySNR}
\end{equation}
In Eq. \ref{equCRdensitySNR} the fraction of the SN explosion energy, 
$E_{SN}={10}^{51}$ ergs, which goes into the CRs is assumed to be $\eta =30 \, \%$. The dependence upon the distance from the 
source, $d$, can be translated into the dependence upon the coordinates, $l$, $b$ and the line of sight distance, $l_d$. The 
minimum proton energy injected is ${E_p}_{min}$ = 1 GeV. In 
Eq. \ref{equCRdensitySNR} $R_d$ is the diffusion distance for a CR of energy $E_p$
\begin{equation}
 R_d = \sqrt{4 \, D(E_p) \, (t-\chi(E_p))}
\label{diffusionDistance}
\end{equation}
where $\chi(E_p)$ is the time after the SN explosion at which a proton of 
energy $E_p$ escapes the injection source and begins diffusing into the ISM. We also assume 
that the CRs released by the SNR have energy dependent diffusion coefficient
\begin{equation}
 D(E_p)  = {D_0 \,\, {[\frac{E_p}{10 \rm{GeV}}]}^{0.5}} \, {\rm{~{cm}^{2} {s}^{-1}} } \, .
\label{equCRdiff}
\end{equation}
We take $D_0$ to be ${10}^{26}$,  ${10}^{27}$ or ${10}^{28}$ cm$^2$/s.
The diffusion of CRs into molecular clouds depends upon the highly
uncertain diffusion coefficient. It is thought that 
CRs can penetrate clouds if the diffusion coefficient inside the cloud is the same as the
average diffusion coefficient in the Galaxy, as derived from spallation
measurements.
CRs are effectively excluded if the diffusion coefficient is suppressed 
compared to the surroundings. 
However, it has been shown that CRs at TeV energies can diffuse into even the densest parts of molecular clouds,
whilst GeV energy CRs might have trouble penetrating the densest parts of molecular clouds 
\citep{Gabici1,Protheroe2008,Jones2009,Gabici2009}. 

\subsection{Hadronic \Gray emissivity}

The hadronic \Gray emissivity, which is given as the \Grays produced by CRs interacting with the gas
per cubic centimeter per second per GeV at a position
defined by its coordinates, $l$, $b$ and heliocentric distance, $l_d$, can be expressed as
\begin{eqnarray*}
&&e_{\gamma}(E_\gamma,l,b,l_d) = \frac{dN_\gamma}{dV dE_\gamma dt } = \\
&&\int_{{E_{p}}_{th}}^{{E_{p}}_{max}} dE_{p} \,\, \frac{d\sigma_{p \rightarrow
\gamma}}{dE_{p}}(E_p,E_\gamma) \,\, \, c  \, \, \, n(l,b,l_d) \, \,
\,\rho_{CR}(E_{p},l,b,l_d)  \,,
\end{eqnarray*}
\begin{equation}
\label{equGammaEmissivity}
\end{equation}
where the integral is calculated over
the energy, $E_{p}$, of the protons. ${{E_{p}}_{th}}$ is the minimum 
proton energy that contributes to the production of a photon of energy $E_\gamma$. In Eq. \ref{equGammaEmissivity} the 
spectral dependence of the photons emitted by CR protons, expressed in terms of the 
differential cross section, 
$d\sigma_{p->\gamma}/dE_{p}$, can be calculated using the simple but precise parametrisation 
developed by \cite{Kelner}. However, at low energies when a broad variety of proton spectra are expected, we 
use the parametrization by \cite{Kamae}. In Eq.\ref{equGammaEmissivity} 
$c$ is the velocity of light and $n(l,b,l_d)$ is the ambient gas density 
as function of $b$, $l$ and $l_d$. $\rho_{CR}(E_p,l,b,l_d)$ 
is the proton energy density, which is given by the sum of the background 
CR density, $\rho_{bg}(E_{p})$, 
and the density of cosmic rays released by the SNR and diffusing into the ISM, 
$\rho_{SNR}(E_{p},l,b,l_d)$, as given in Eq.\ref{equCRdensitySNR}
\begin{equation}
\rho_{CR}(E_{p},l,b,l_d) = \rho_{bg}(E_{p}) + \rho_{SNR}(E_{p},l,b,l_d) \,.
\label{equcrdensity}
\end{equation}
The CR background spectrum is equal to the CR flux measured at the top of 
the Earth's atmosphere. Below ${10}^6$ GeV we here adopt the value of
\begin{equation}
 \Phi_{bg}(E_{p}) = 1.8 \, \, {( \frac{E_{p}}{\rm{GeV}})}^{-2.7} \, {\rm {cm}^{-2}}{\rm s^{-1}}{\rm {sr}^{-1}}{\rm {GeV}^{-1}} \,,
\label{eqnSEA}
\end{equation}
taken from the \cite{ParticleDataGroup}. A change in the spectral index of 
the measured CR flux is observed above ${10}^6$ GeV, which we do not consider here since 
we are only interested in the \Gray emission below 100 TeV. The background CR energy density is then defined as
\begin{eqnarray*}
\rho_{bg}(E_{p}) &=&   \frac{4 \pi}{c} \, \Phi_{bg}(E_{p}) \\
&\simeq& 7.5 \times {10}^{-10} \, { \frac{E_{p}}{\rm {GeV}}}^{-2.7} \, {\rm {cm}^{-3}} {\rm {GeV}^{-1}} \,.
\end{eqnarray*}
\begin{equation}
\label{equcrSEAdensity}
\end{equation}
A multiplication factor of 1.5 is applied to the CR density in Eq.\ref{equcrdensity}, 
which accounts for the contribution to the emission from heavier nuclei both in
CRs and in the interstellar medium \citep{Dermer,Mori}. We note that recently 
\cite{Mori2009} has recalculated the multiplication factor obtaining a 20 per cent higher value.

The differential photon spectrum measured at Earth is then obtained by
integrating the emissivity
$e_{\gamma}(E_\gamma,l,b,l_d)$ over the line of sight distance $l_d$
\begin{equation}
\frac{dN_\gamma}{dA dE_\gamma dt d\Omega } = \frac{1}{4 \pi}\,
\int_{0}^{{l_d}_{max}} d l_d \, e_\gamma(E_\gamma,l,b,l_d)
\label{equGammaFlux}
\end{equation}

\subsection{Leptonic emission from primary and secondary electrons}

\Grays are also produced through inverse Compton and 
bremsstrahlung processes of very highly energetic primary and secondary electrons and positrons. 

A population of primary electrons is accelerated by and confined within the SNR. 
The high energy primary electrons suffer, in fact, from strong losses 
through synchrotron emission, triggered by the amplified magnetic field of the SNR shock, 
while the low energy part of the primary leptonic population cannot escape due to diffusive confinement.

Inverse Compton (IC) scattering of background CR electrons results in a
non-negligible contribution to the overall diffuse gamma-ray emission at
GeV, and also at TeV energies, especially 
at high Galactic latitudes. At low latitudes the relative contribution of this component, compared to
the bremsstrahlung and $\pi^0$-decay gamma-rays from specific dense 
regions is significantly reduced because of the enhanced gas density. 
Following 
\cite{Aharonian2000b} in Fig.\ref{fig66} we show the contribution to 
the emission due to IC scattering of background electrons to the emission spectra.
The ratio of luminosities from non-thermal bremsstrahlung 
to hadronic $\gamma$-rays does not depend on the ambient gas
density. However, the contribution of the bremsstrahlung of background electrons 
to the total diffuse emission above 1 GeV is rather low, so it can be ignored.

When CRs interact with ambient matter through inelastic collisions
they produce not only neutral pions, but also charged pions. These charged
pions decay into secondary electrons, positrons and neutrinos ( see \cite{Gabici2009}).
Thus, in addition to the \Gray emissivity due to neutral pion decay,
there will be a simultaneous radio synchrotron component due to the secondary
electrons and positrons produced concomitantly with the neutral pions. 
The secondary leptons will also produce \Gray emission through
bremsstrahlung and inverse Compton. However,
as discussed in \cite{Gabici2009}, the contribution of bremsstrahlung and
inverse
Compton of secondary leptons is negligable and the pion decay emission is 
dominant in the large energy range from GeV to TeV energies,
assuming for the protons the energy spectrum given by Eq.\ref{eqnSEA}. 
The radio emission of primary and secondary electrons in the enviroments 
surrounding SNRs is of interest, but is beyond the scope of this paper and will not be 
discussed here.

\section{Results and discussion}\label{sec:results}

\subsection{The emission of background CRs}

\begin{figure}
\begin{centering}
\includegraphics[width=0.35\textwidth,angle=0]{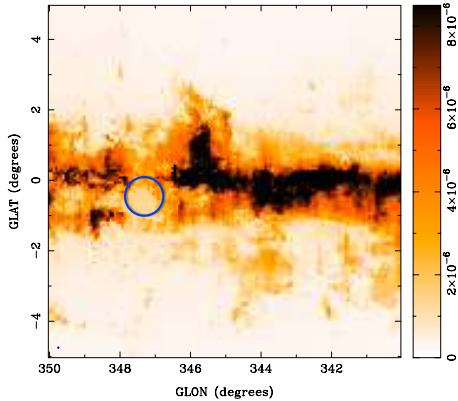}
\caption{The $\gamma$-ray energy flux arising from the background CRs,
expressed in GeV cm$^{-2}$ sr$^{-1}$ s$^{-1}$ at 1 TeV.
The blue ring represents a sphere of 10 pc radius around the
location of the SNR RX~J1713.7-3946
(the radial extension of the radio SNR shell).\label{fig1}}
\end{centering}
\end{figure}

We first calculate the spatial features 
of the hadronic \Gray emission under the assumption that the CR spectrum in the region 
of longitude range $340^\circ<l<350^\circ$ and latitude range 
$-5^\circ<b<5^\circ$ is uniform and equal to the locally observed CR flux.
Figure \ref{fig1} shows the $\gamma$-ray emission from this region arising from the 
interactions of background CRs, whose spectrum is given in Eq.\ref{eqnSEA}, 
with the ambient atomic and molecular hydrogen, as measured by the LAB and NANTEN surveys.
The background CR spectrum being uniform, 
the morphology of the emission reproduces the gas distribution and does
not change with energy. In Figure \ref{fig1} the emission at 1 TeV is plotted.
The blue ring represents a sphere of 10 pc radius (the radial extension of 
the radio SNR shell) 
around the location of the SNR RX~J1713.7-3946. Hereafter we assume 
that the CR spectrum within 10 pc radius from RX~J1713.7-3946 is equal to 
the CR background spectrum, since we are interested in the
emission of CRs leaving their injection source and not in the emission
produced by the injected CRs within the SNR shell. 

\subsection{The emission of background and runaway CRs}

\begin{figure}
\begin{centering}
  \includegraphics[width=0.31\textwidth,angle=-90]{figure3.eps}\\
  \includegraphics[width=0.35\textwidth,angle=90]{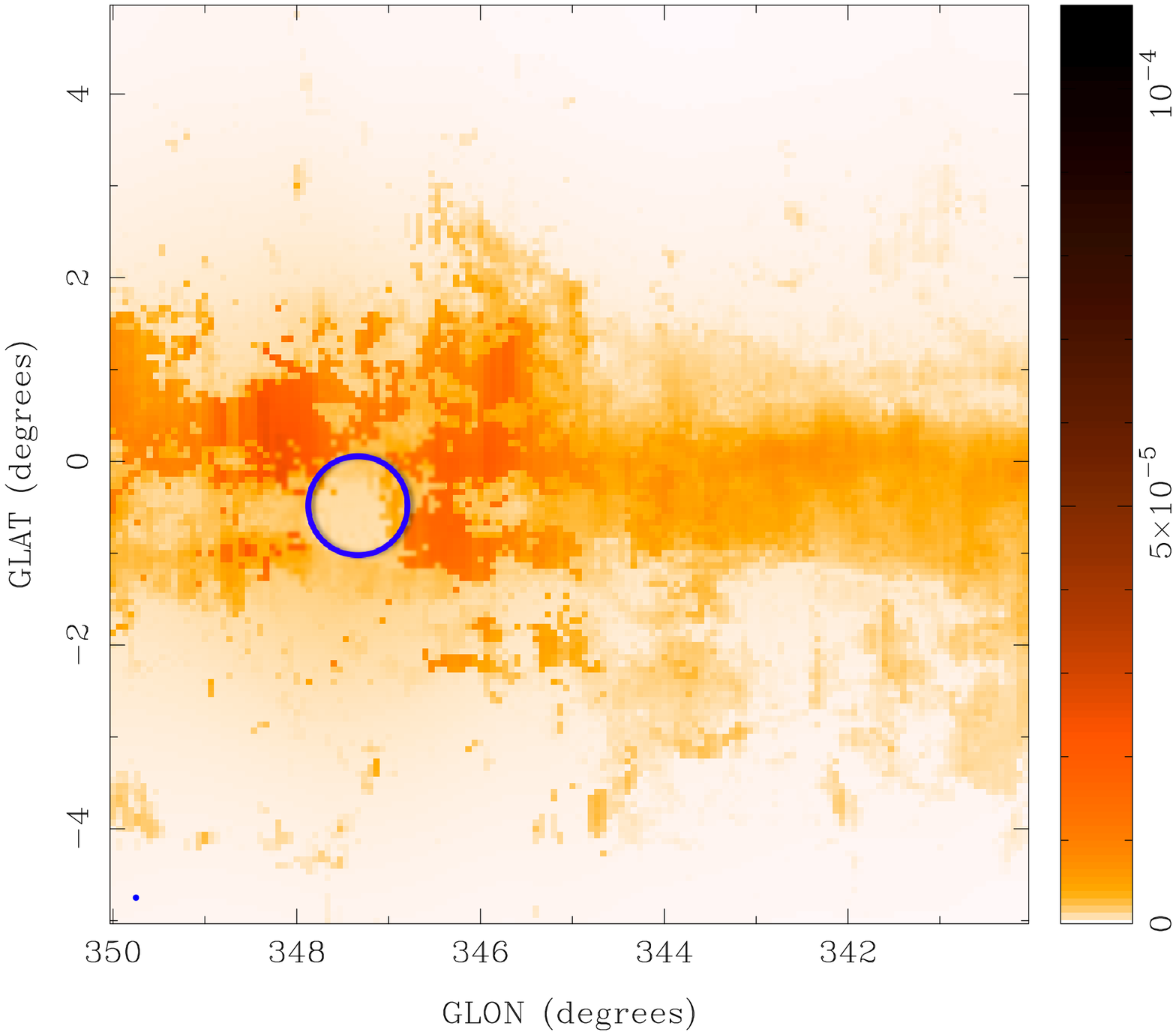}\\
  \includegraphics[width=0.35\textwidth,angle=-270]{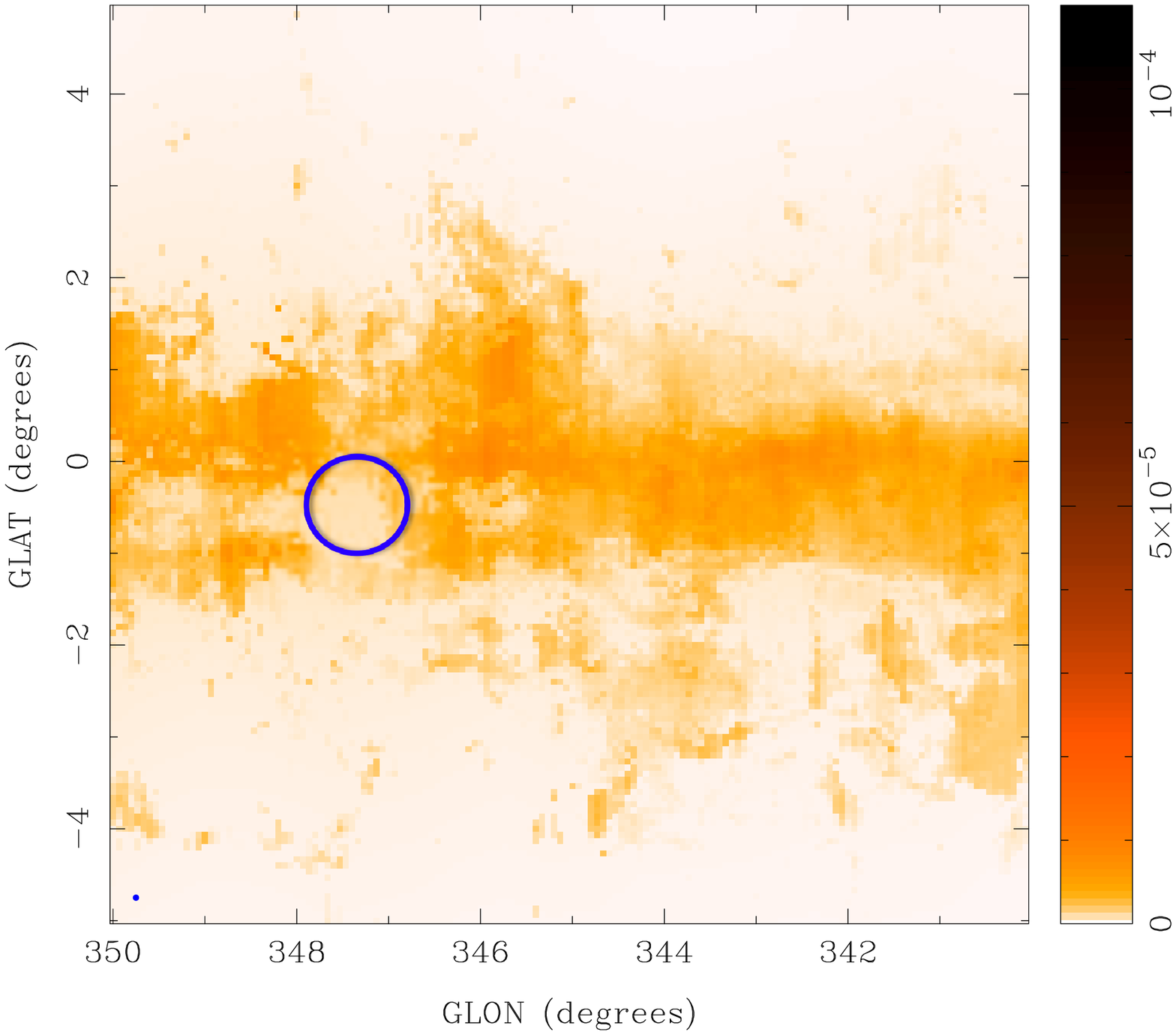}\\
\caption{ The predicted hadronic $\gamma$-ray energy fluxes at 1 TeV, expressed in GeV
cm$^{-2}$ sr$^{-1}$ s$^{-1}$, arising from background CRs and from CRs escaping the SNR shells
if the CR diffusion coefficient, $D_0$, is equal to 10$^{26}$ cm$^2$/s (upper 
panel), 10$^{27}$ cm$^2$/s (middle panel) and 10$^{28}$ cm$^2$/s (bottom panel). We assume that the SNR has
started releasing the particles of energy 500 TeV 100 years after the explosion. 
Also indicated in the left bottom corner the
angular resolution of the NANTEN survey.\label{fig5}}
\end{centering}
\end{figure}

Figure \ref{fig5} shows the predicted hadronic \Gray emission at 1 TeV 
arising from the sum of background CRs and the runaway CRs (i.e., ${\rho}_{bg} + {\rho}_{SNR}$) as given by Eq. \ref{equcrdensity} 
for the diffusion coefficients discussed in Section~3. The spatial distribution of the emission depends 
upon the CR diffusion coefficient, $D_0$. The faster the CRs diffuse 
into the ISM, the further the enhanced emission extends beyond the linear extent of RX~J1713.7-3946. However, 
for very fast diffusion ($D_0= 10^{28}$ cm$^2$/s) the VHE runaway protons have already left the region around the SNR. 
The shape of the emission depends also upon the assumptions concerning the SNR injection
history. We have namely assumed that the
SNR, which is now about 1600 years old, has started to inject the most
energetic particles at $t_{Sedov}$=100 years after the explosion.
According to Eq.\ref{eqn:chi} this assumption means that nowadays the SNR starts releasing CRs of energy about 100 TeV.
CRs of energy lower than 100 TeV are still confined
in the SNR shell, whereas higher energy particles have already left it. 
The \Gray emission produced by a proton of 100 TeV
peaks at around 10 TeV. Roughly speaking, CRs of a given energy are responsible for hadronic \Gray 
emission which is about 10 times less energetic. The assumption that 
CRs of energy below 100 TeV are still confined
in the SNR shell means then that one expects to observe almost no \Gray enhanced emission 
due to runaway protons at energies below few TeVs in the surroundings of the SNR RX~J1713.7-3946.
The assumption on the injection history 
of the CR source is therefore crucial to determine the spatial and spectral features of the \Gray emission. 

One would expect that the \Gray emission from MCs illuminated by the CRs
injected from nearby CR sources
is spatially correlated with the atomic and molecular gas distribution.
This is, generally speaking, true.
However, one has to take into account that both the acceleration history of the
source and the
diffusion timescale are processes dependent upon the energy. In
other words, when considering
\Gray emission at a given energy from the region surrounding a SNR, one should 
expect a correlation
between hadronic \Gray and the gas distribution if and only if the parent CRs have already been
released by the SNR
and had time enough to diffuse into the ISM. On the other hand, the
enviroment of a CR source, which is dense
in atomic and molecular hydrogen, might nonetheless appear faint at some
energies in \Grays simply because the parent CRs have not escaped the injection source or 
have not had time enough to propagate throughout the region nearby the source. 

\begin{figure}
\centering
  \includegraphics[width=0.22\textwidth]{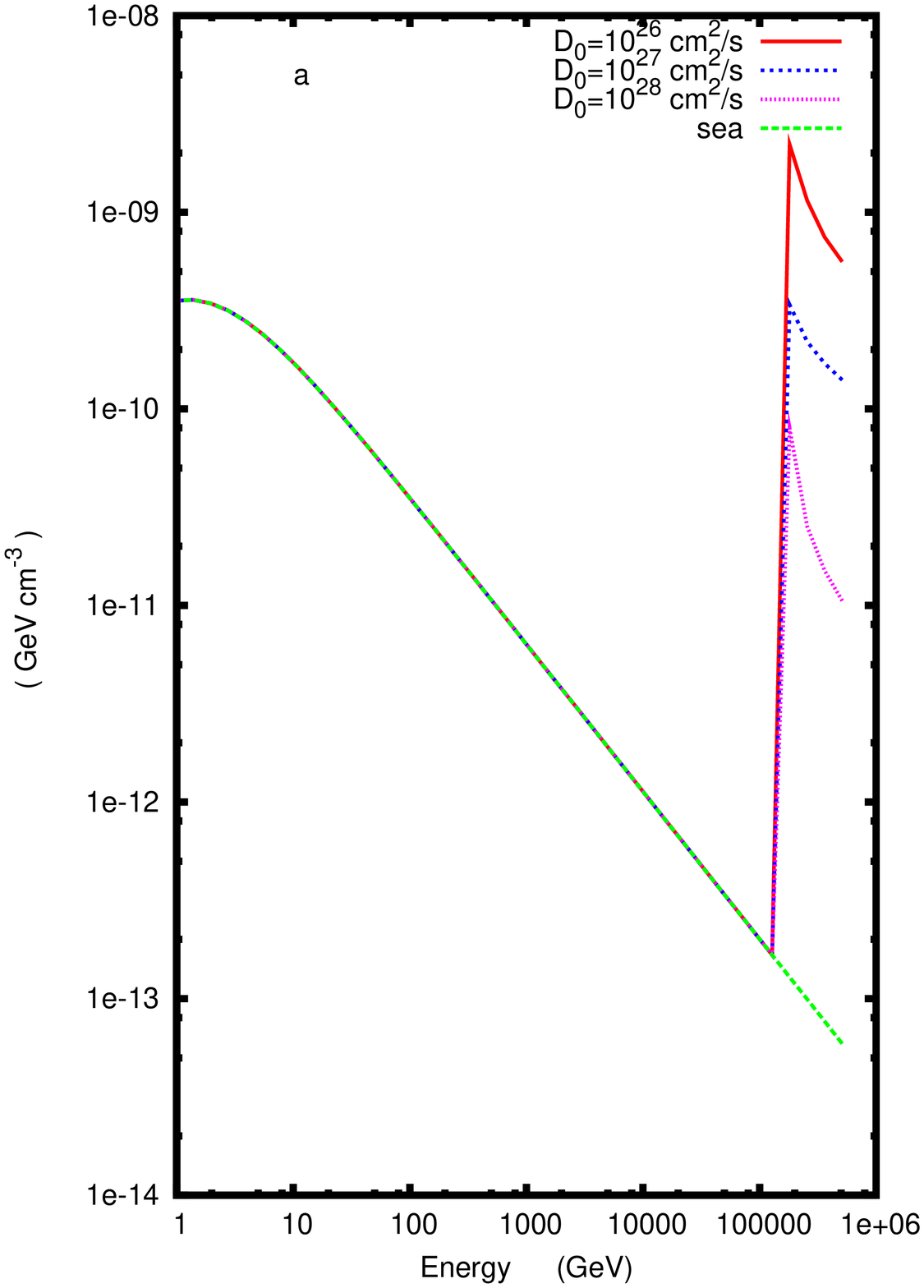}
  \includegraphics[width=0.22\textwidth]{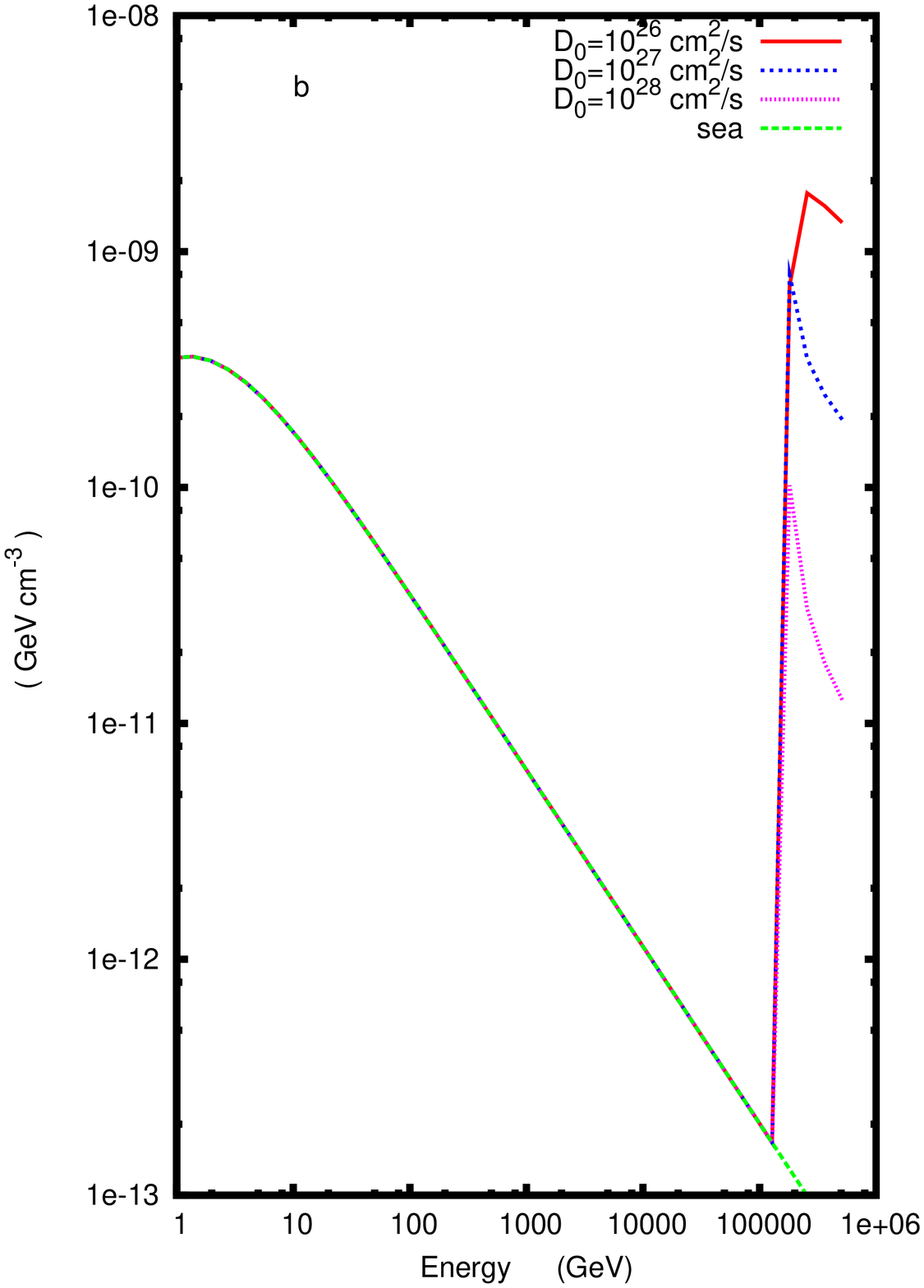}
  \includegraphics[width=0.22\textwidth]{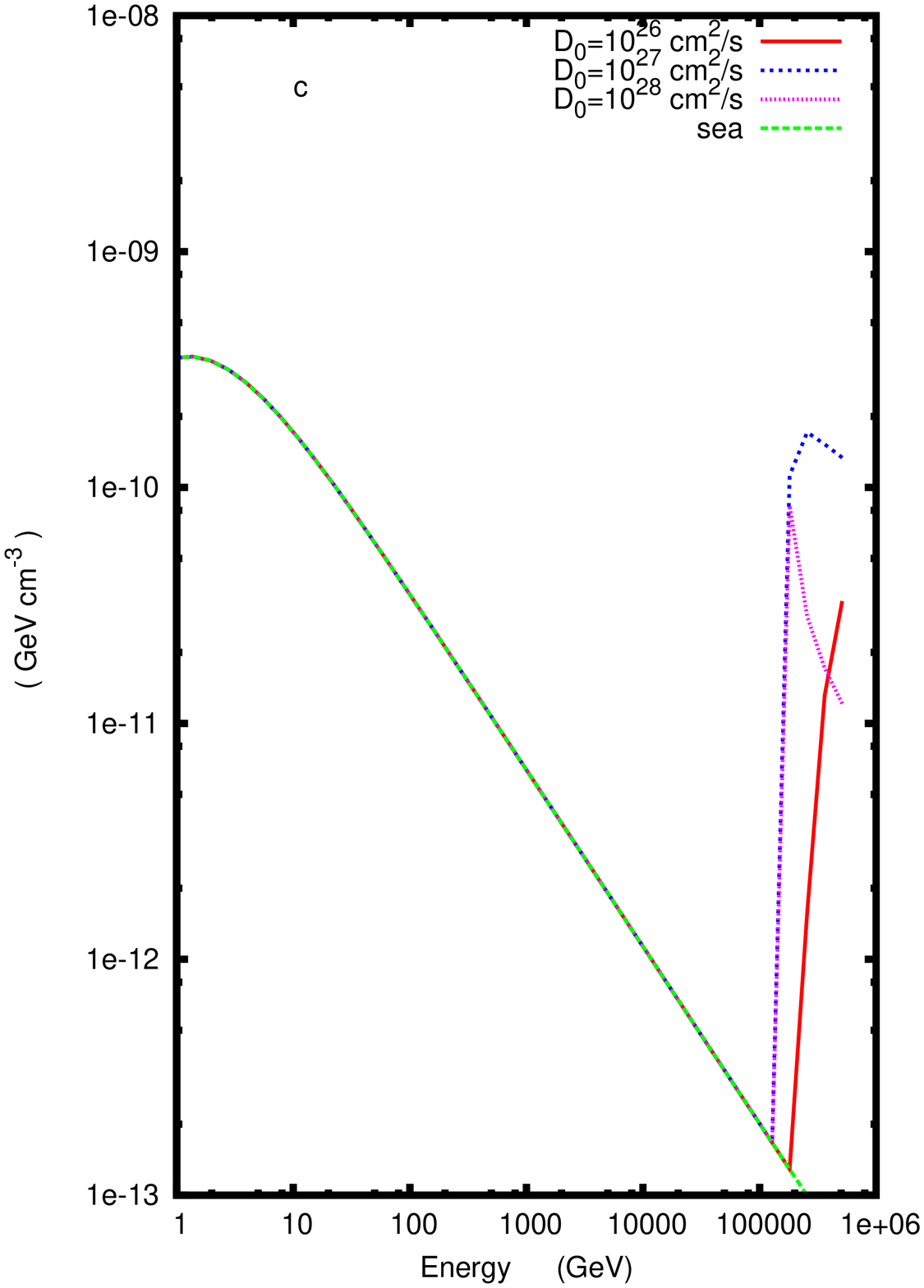}
  \includegraphics[width=0.22\textwidth]{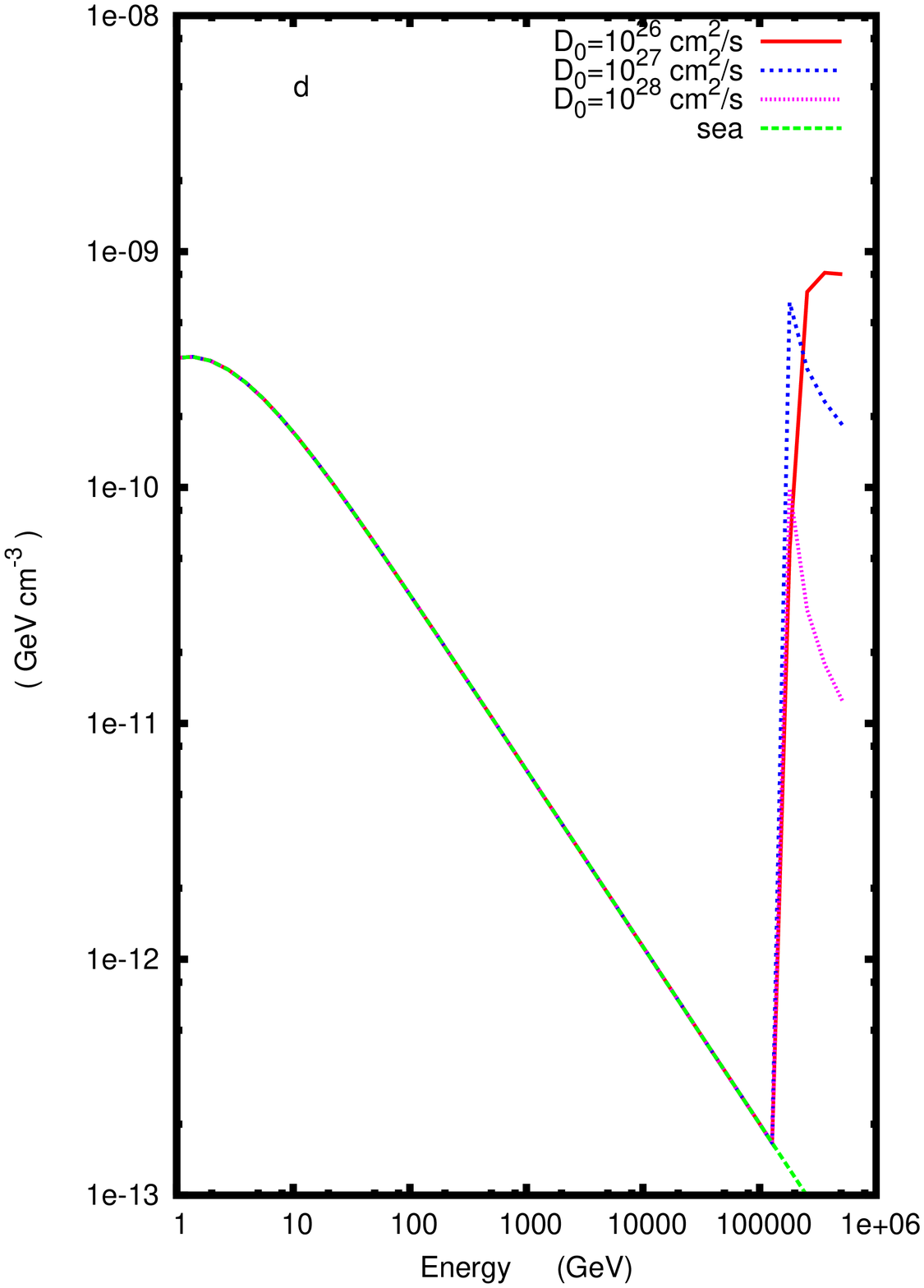}
\caption{The average CR proton energy density in four different regions 
of 0.2 $\times$ 0.2 degrees around the positions a = (346.8, -0.4), b = (346.9, -1.4), c = (347.1, -3.0) and d = (346.2, 0.2). 
The CR density is averaged over 200 parsecs around 1 kpc along 
the line of sight distance. The CR energy distributions in the panels a,b,c and d, corresponding to the different locations, 
are plotted for different diffusion coefficients $D_0$. The background CR distribution is also shown in each panel for comparison.\label{fig55}}
\end{figure}

Figure \ref{fig55} shows the average CR energy density from four different regions of 0.2 $\times$ 0.2 degrees 
around the positions a = (346.8, -0.4), b = (346.9, -1.4), c = (347.1, -3.0) and d = (346.2, 0.2). 
The CR energy density from these four regions is averaged over 200 parsecs around 1 kpc along 
the line of sight distance. If averaged over the whole line of sight distance from 50 parsecs to 30000 parsecs the CR energy density 
would be dominated by the energy density of the background. The background 
CR energy distribution is also shown in each panel a, b, c and d of Figure \ref{fig55} for comparison. 
The CR energy distributions in the panels a, b, c and d, corresponding to the different locations and 
plotted for different diffusion coefficients $D_0$, vary depending upon the location and the diffusion coefficient. 
The average gas density in the four regions a,b,c and d is about 3, 1, 1 and 8 protons/cm$^3$, respectively. 
The total mass in solar masses, $M_{\sun}$, contained in these four regions is 458, 100, 88 and 1013 $M_{\sun}$ 
for distances between 900 and 1100 parsecs. 

\begin{figure}
\centering
  \includegraphics[width=0.22\textwidth]{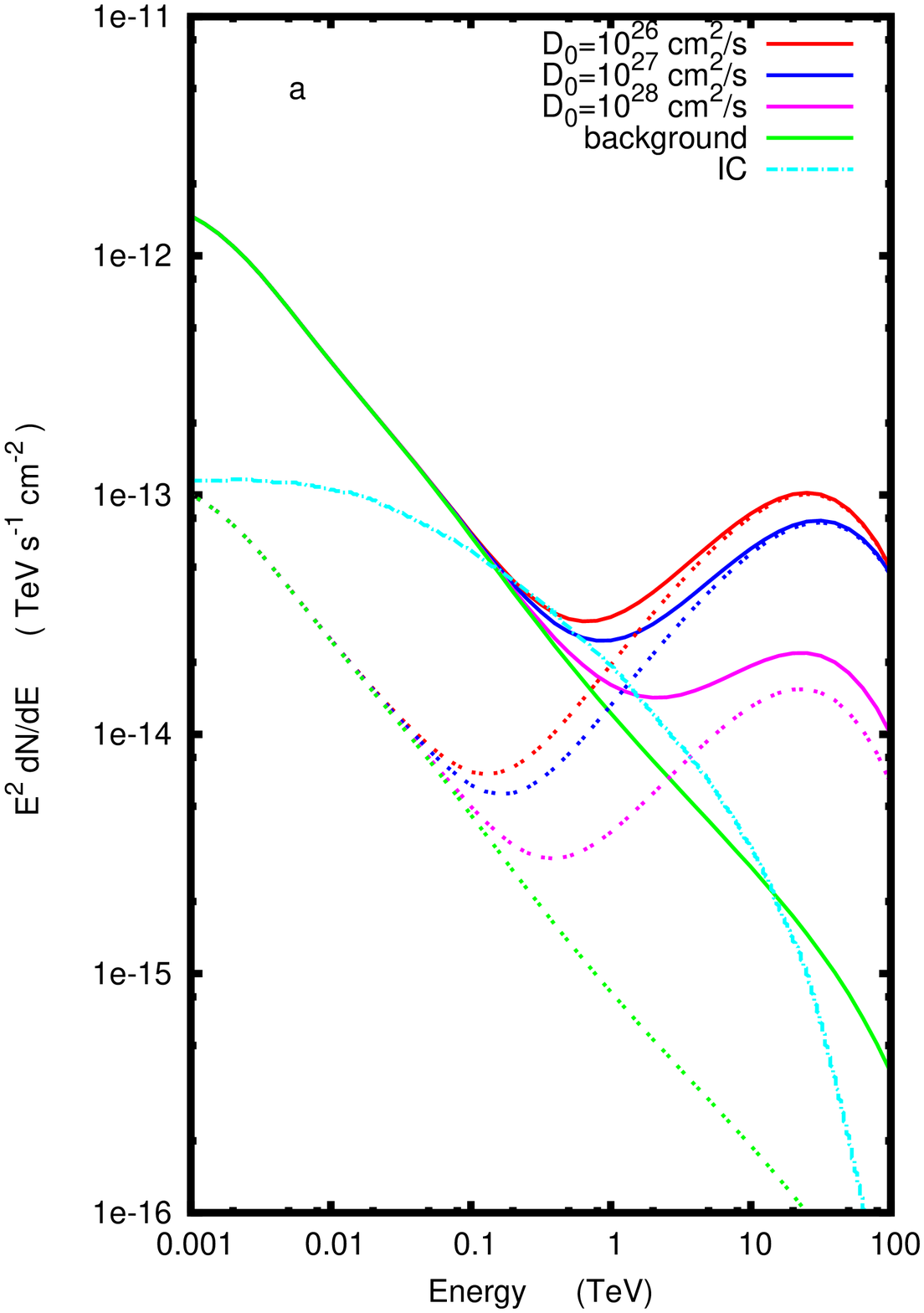}
  \includegraphics[width=0.22\textwidth]{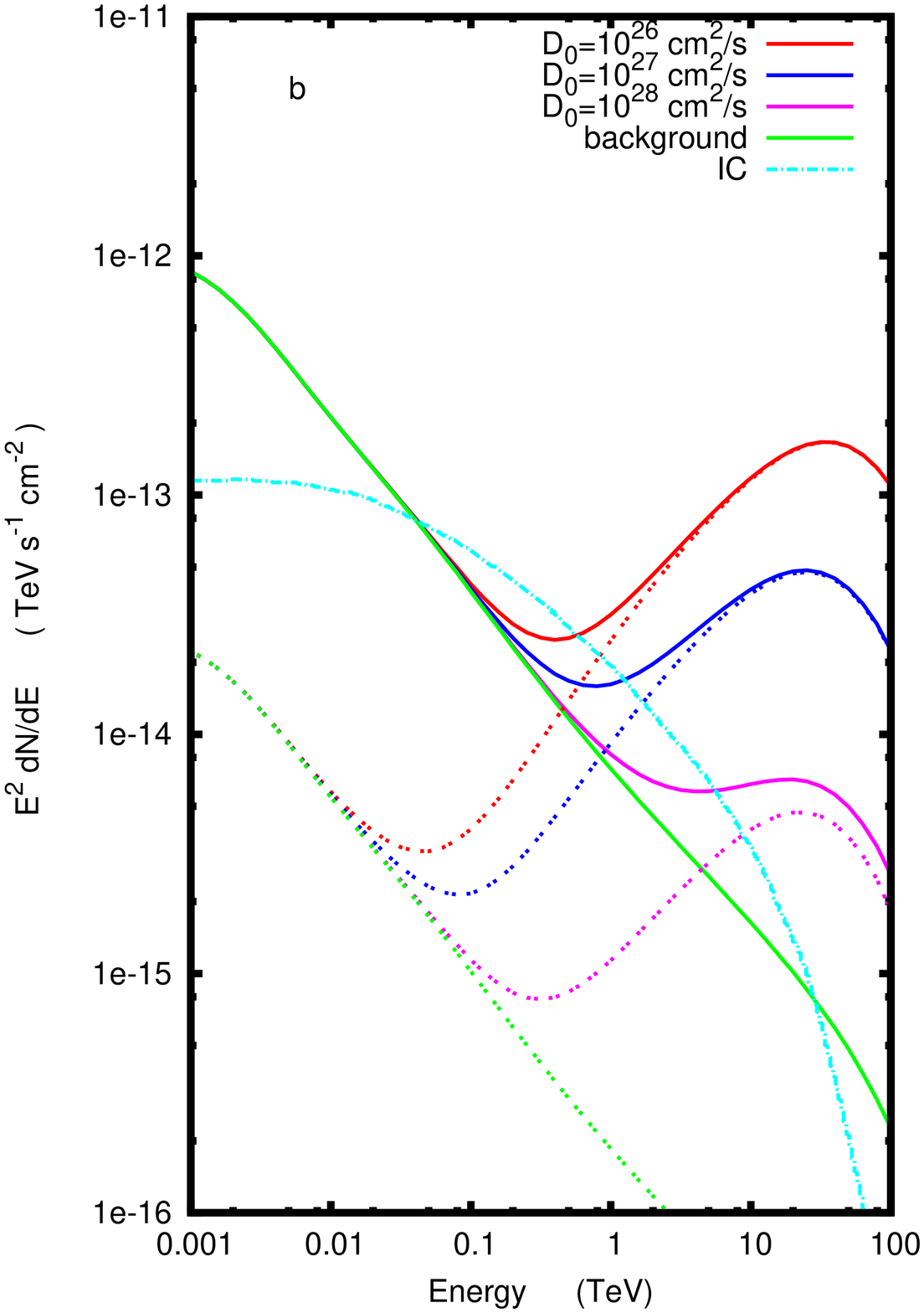}
  \includegraphics[width=0.22\textwidth]{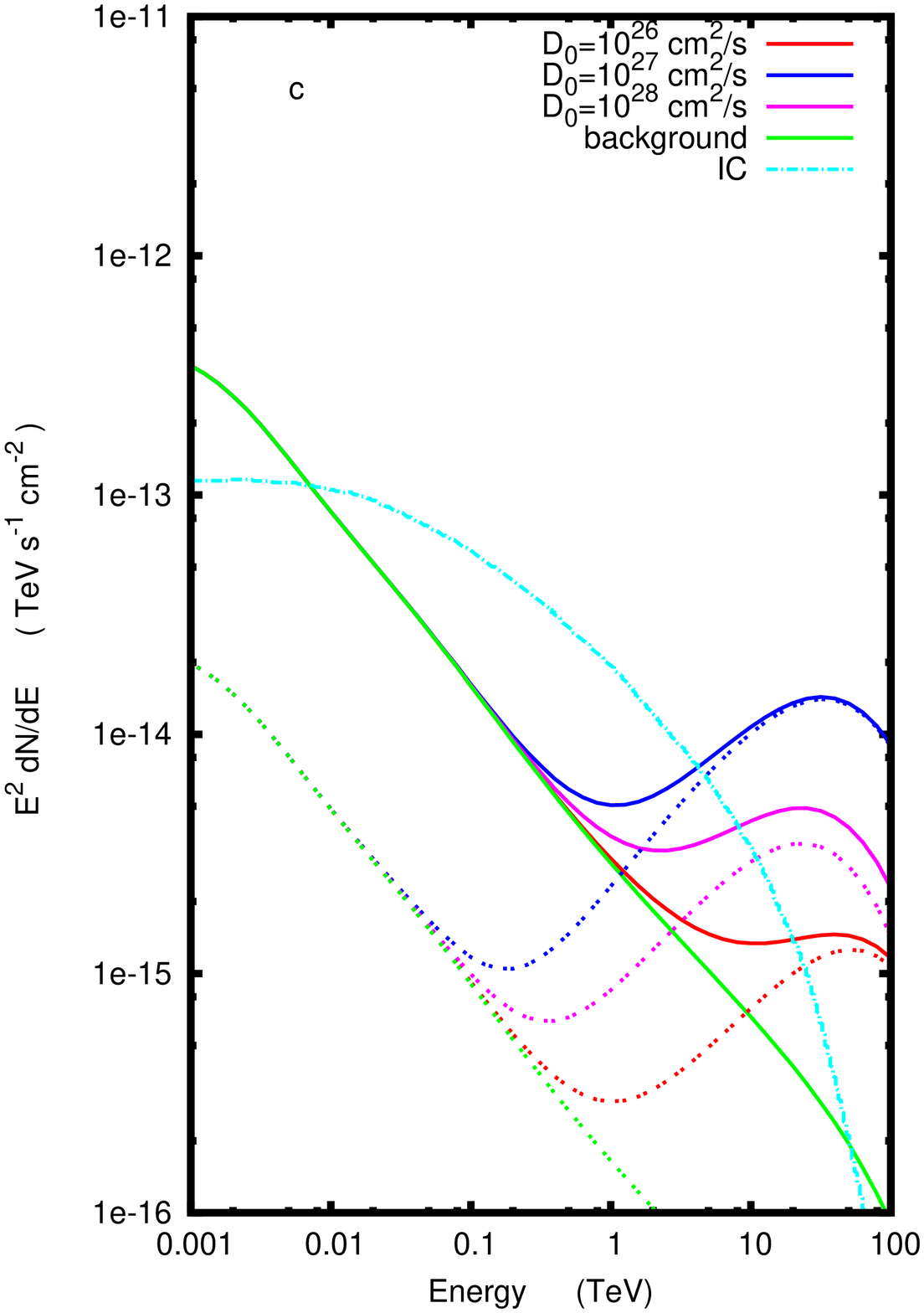}
  \includegraphics[width=0.22\textwidth]{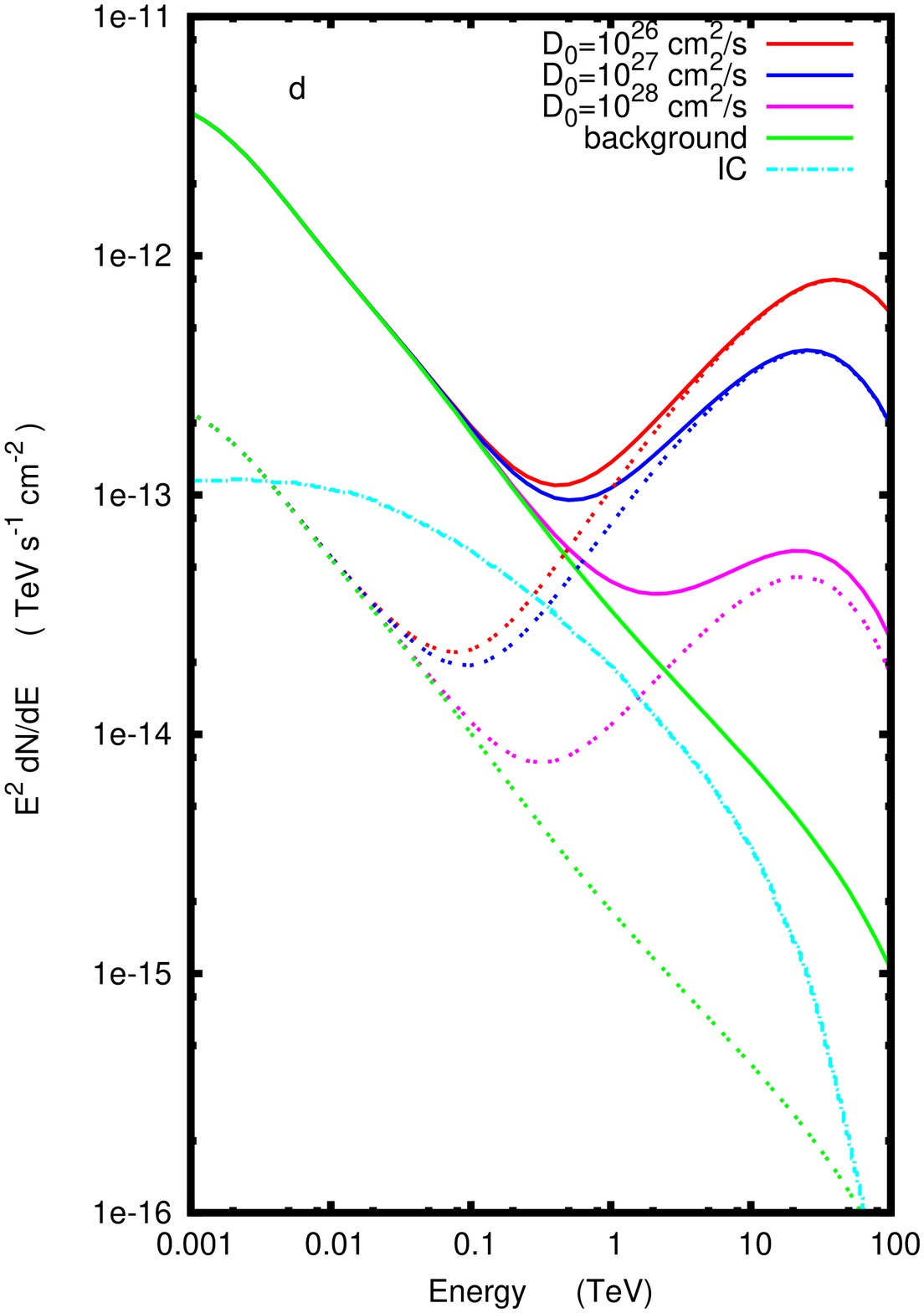}
\caption{The $\gamma$-ray energy flux in four different regions of 0.2 $\times$ 0.2 degrees 
around the positions a = (346.8, -0.4), b = (346.9, -1.4), c = (347.1, -3.0) and d = (346.2, 0.2). 
The emission produced along the line of sight distance between 
900 and 1100 parsecs is plotted in dashed lines, while the emission 
obtained by summing the radiation contributions over the whole line of sight distance, from 50 parsecs to 30000 parsecs, is 
shown in solid lines. The emission in the panels a,b,c and d, corresponding to the different locations, is 
plotted for different diffusion coefficients $D_0$. The emission from background 
CRs is also shown in each panel for comparison. The contribution to the emission from inverse Compton scattering of 
background electrons is 
indicated with a dashed light blue line. For the modeling of the inverse Compton contribution we follow \cite{Aharonian2000b}. \label{fig66}}
\end{figure}

Figure \ref{fig66} shows the $\gamma$-ray spectra from regions a,b,c and d 
if one considers the hadronic emission produced along the line of sight distance between 900 and 1100 parsecs 
(in dashed lines) and the hadronic emission 
obtained by summing the radiation contributions over the whole line of sight distance (in solid lines). 
The contribution to the emission from inverse Compton scattering of 
background electrons, obtained following the modeling of \cite{Aharonian2000b}, 
is indicated with a dashed light blue line. In all four locations the  hadronic $\gamma$-ray emission is enhanced with respect 
to the hadronic emission due only to background CRs at energies above few TeVs as a consequence 
of the fact that the CR fluxes are enhanced above 100 TeV. The hadronic $\gamma$-ray emission 
from the region d is particularly enhanced with 
respect to the hadronic emission from background CRs. This is due to 
two factors, the enhanced CR flux nearby the injection source and the high ambient gas density. 
In the three regions a, b and d 
the high energy emission is clearly dominated by the radiation produced along the line of sight 
distance between 900 and 1100 parsecs, plotted 
in dashed lines in Figure \ref{fig66}. The $\gamma$-ray emission from high latitude regions, such as 
region c, is instead dominated by the contribution from IC scattering of background electrons, 
almost at all energies. 
In regions closer to the Galactic Plane the emission from inverse Compton scattering of background electrons 
is subdominant at TeV energies, where runaway cosmic rays produce the enhanced emission. Therefore the regions 
where to look for the emission from runaway particles are low latitude regions of higher gas density.

The $\gamma$-ray spectra show a peculiar concave shape, being soft at low energies and hard at high energies, which, 
as discussed in \cite{Gabici2009}, might be important for the studies of the spectral compatibility 
of GeV and TeV gamma ray sources.  The peculiar spectral 
and morphological features of the \Gray due to runaway CRs can be 
therefore revealed by combining the spectra and \Gray images provided by the Fermi and Agile telescopes 
at GeV and by present and future ground based detectors at TeV energies. 
As shown by the surveys of the Galaxy, published by Fermi 
at MeV-GeV energies \citep{Abdo:2009}, by HESS at TeV energies \citep{Aharonian2005,Aharonian2006} 
and at very high energies by the Milagro Collaboration \citep{Abdo2007,Milagro2009}, the 
various extended Galactic sources differ in spectra, flux and morphology. However, there is growing 
evidence for the correlation of GeV and TeV energy sources \citep{Funk,Milagro2009}. These 
sources appear often spectrally and morphologically different at different energies, possibly 
due not only to the better angular resolution obtained by the instruments at TeV energies, 
but also to the energy dependence of physical processes, such as CR injection and CR 
diffusion. For this reason it is important to properly model what we expect to observe at different energies 
by conveying in a quantitave way all information by recognizing that the enviroment, 
the source age, the acceleration rate and history, all play a role in the physical 
process of injection and all have to be taken into account for the predictions. 

\begin{figure}
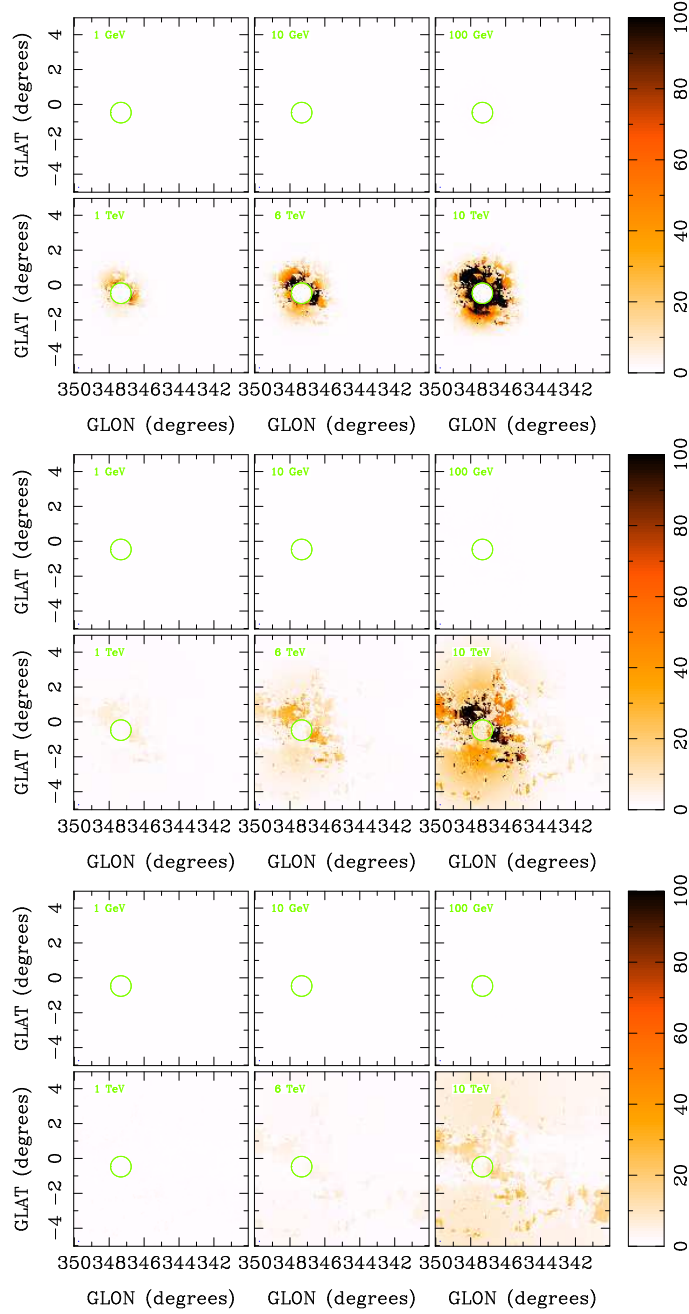

\centering
\includegraphics[width=0.25\textheight,angle=-90]{figure8.eps}\\
\includegraphics[width=0.25\textheight,angle=-90]{figure9.eps}\\
\includegraphics[width=0.25\textheight,angle=-90]{figure10.eps}\\
\caption{Ratio of the emission due to the sum of background CRs and runaway CRs 
and background CRs only. The SNR is supposed to have exploded at 1 kpc distance from the Sun at ${347.3}^{\rm o}$
longitude and $-0.5^{\rm o}$ latitude 1600 years ago and to have started injecting the 
most energetic protons 100 years after the explosion .
The diffusion coefficient assumed within the region $340^{\rm
o}<l<350^{\rm o}$ and  $-5^{\rm o}<b<5^{\rm o}$ is 10$^{26}$ cm$^2$/s in
the upper panel, 10$^{27}$ cm$^2$/s in the middle panel and 10$^{28}$ cm$^2$/s in
the bottom panel. In the three panels the ratio of the emission is shown for different energies 
from 1 GeV to 10 TeV.}\label{fig2}
\end{figure}

Figure \ref{fig2} shows the ratio of the hadronic gamma-ray emission due to total CR spectrum to that of the
background CRs for the entire region under consideration. In our modeling only CRs with energies above about 100 
TeV have left the acceleration site and the morphology of the emission depends upon the 
energy at which one observes the hadronic gamma-ray emission. The different 
spatial distribution of the emission is also due to the different energy-dependent 
diffusion coefficients, assumed in the three different panels. 

As discussed at length in Paper 1 the distance to the molecular and atomic gas is 
highly uncertain. The principle uncertainty in the determination of the
distance, which especially at distances close to the Sun, is as large as 2~kpc,  
comes from errors in the accuracy of the estimates of radial velocity of gas clouds. 
Uncertainties in the distance estimates affect also the determination 
of the conversion factor X in Eq. \ref{eq:X} \citep{arimoto}, by which the emissivity of the CO line 
is converted to ambient matter density, and the determination of the number  
density of H$_2$ molecules from the measured CO intensity. As noted in Paper 1, these uncertainties do 
not affect the model predictions of the hadronic gamma-ray flux. This is 
because the hadronic gamma-ray emissivity is proportional to the gas 
mass and inversely proportional to the distance squared so that the relative
errors exactly cancel.

\section{Conclusions}\label{sec:conclusions}

CRs escaping SNRs diffuse into the ISM and collide with the ambient atomic and molecular gas. 
From such collisions $\gamma$-rays are created, which can possibly provide 
the first evidence of the parent population of runaway CRs. 
The \Gray radiation from such hadronic interactions from regions close to CR sources depends not only on the total
power emitted in CRs by the sources, and on the distance 
of the source to us, but also on the ambient interstellar
gas density, the local diffusion coefficient and the injection history of the CR source. 
Due to the fact that the emission from gas clouds surrounding SNRs, illuminated in gamma-rays, 
depends upon many factors, it is difficult to draw strong conclusions about the nature of such emission. 
In order to test the standard scenario for CR 
injection in SNRs and CR propagation, we have computed the expected hadronic gamma-ray 
emissivity for the region $340^\circ<l<350^\circ$ and $-5^\circ<b<5^\circ$, 
assuming that a SN event has occurred in the
location of the historical SNR RX~J1713.7-3946 in 393 C.E.. Detailed modeling of the 
energy spectra and of the spatial distribution of the gamma-ray emission
using the data from atomic and molecular hydrogen in the environment surrounding 
RX~J1713.7-3946 have been presented. These predictions have shown that 
the age and acceleration history of the SNR, the particle 
diffusion regime and the distribution of the ambient gas are all 
paramount. The emission from the regions surrounding SNR shells can therefore provide crucial informations on the
history of the SNR acting as a CR source and important constraints on the highly unknown diffusion coefficient.
Also, depending on the time and energy at which one observes 
the remnants and the surrounds, one will observe different spectra and morphologies. 
This has important implications for the current and future generations of gamma-ray observatories. The 
high sensitivity and high resolution, which will be reached 
by future detectors, such as AGIS, CTA and HAWC (e.g. see reviews \citep{Sinnis,Hinton}), makes the detection of 
the predicted emission possible.

\section{Acknowledgements}

Sabrina Casanova acknowledges the support from the 
European Union under Marie Curie Intra-European fellowships.
The NANTEN telescope was operated based on a mutual agreement between 
Nagoya University and the Carnegie Institution of Washington.
We also  acknowledge that the operation of NANTEN was realized by
contributions  from many Japanese public donators and companies.
This work is financially supported in part by a Grant-in-Aid for
Scientific Research from the Ministry of Education, Culture, Sports,
Science and  Technology of Japan (Nos. 15071203 and 18026004, and
core-to-core  program No. 17004), JSPS (Nos. 14102003, 20244014, 18684003, and 22540250),
and the Mitsubishi foundation.

-------------------------------------------------------------------

\end{document}